\documentstyle[axodraw]{article}

\title{
\begin{flushleft}{\normalsize hep-ph/9712414 \\
SINP/TNP/97-23}
\end{flushleft}
Optical activity of a neutrino gas}
 
\author{
Subhendra Mohanty\\ 
Physical Research Laboratory \\
Navrangpura, Ahmedabad 380009, INDIA \\ 
\and
Jos\'e  F. Nieves\\
Laboratory of Theoretical Physics \\
Department of Physics, University of Puerto Rico\\
P.O. Box 23343, Rio Piedras, Puerto Rico 00931-3343 \\ 
\and
Palash B. Pal \\
Theory Group, Saha Institute of Nuclear Physics \\
Block-AF, Bidhan-Nagar, Calcutta 700064, INDIA}
 
\date{December 1997}

\begin{document}

\maketitle

\begin{abstract} 

For photons that propagate through a gas of 
neutrinos with a non-zero chemical potential,
the left-handed and right-handed polarization modes
acquire different dispersion relations.  This is due to the
$CP$ and $CPT$-odd terms induced by such a background
on the photon self-energy.
We present a detailed calculation of this effect,
which does not depend on any physical assumptions
beyond those of the standard electroweak model.
Some possible cosmological and astrophysical implications
of our results are considered in several contexts, including  the recent
discussions regarding the rotation of the plane of
polarization of electromagnetic waves over cosmological distances.

\end{abstract}

\section{Introduction}
\setcounter{equation}{0}
It was shown some time ago by two of the present authors that, in 
an isotropic medium, the most general parameterization of the
photon polarization tensor $\Pi_{\mu\nu}$ admits three invariant 
form factors. This is due to the fact that, in addition
to the the photon momentum $k^\mu$, $\Pi_{\mu\nu}$
depends also on the velocity four vector of the 
center of mass of the background medium, which we denote by $u^\mu$. 
Thus, the most general form of $\Pi_{\mu\nu}$ consistent with the
gauge invariance conditions
	\begin{eqnarray}
k^\mu \Pi_{\mu\nu} (k) = 0 \,, \qquad
k^\nu \Pi_{\mu\nu} (k) = 0 \,,
	\end{eqnarray}
is 
	\begin{eqnarray}
\Pi_{\mu\nu} (k) = \Pi_T R_{\mu\nu} + \Pi_L Q_{\mu\nu} + \Pi_P
P_{\mu\nu} \,,
\label{Pi}
	\end{eqnarray}
where 
	\begin{eqnarray}
R_{\mu\nu} &=& g_{\mu\nu} - {k_\mu k_\nu \over k^2} - Q_{\mu\nu}
\,, \\* 
Q_{\mu\nu} &=& {\widetilde u_\mu \widetilde u_\nu \over \widetilde
u^2} \,, \\* 
P_{\mu\nu} &=& {i\over K} \varepsilon_{\mu\nu\lambda\rho} k^\lambda
u^\rho \,,
\label{PQR}
	\end{eqnarray}
with
	\begin{eqnarray}
K \equiv \sqrt{\omega^2 - k^2} \,, \qquad \omega \equiv k \cdot u
\,.
\label{K}
	\end{eqnarray}

In Eq.\ (\ref{Pi}), $\Pi_{T,L,P}$ are scalar functions of the Lorentz
invariants $K$ and $\omega$, which have the interpretation of being
the momentum and energy of the photon in the frame in which the medium
is at rest. The functions $\Pi_{T,L}$ are related to the well-known
dielectric and magnetic permeability functions of the medium
\cite{epsmu}, while $\Pi_P$ is related to a third constant that is
responsible for natural optical activity. This third constant, which
was called the {\em activity constant}\/ in Ref.~\cite{activity}, can
arise only as a result of parity ($P$), $CP$ as well as $CPT$
asymmetric effects. This does not necessarily mean that
these symmetries have to be violated at a fundamental level. The
desired asymmetries may also come from the background medium. In
particular, the $P$ and $CP$ breaking effects may arise either from
the violation of these symmetries at the level of the fundamental
interactions, or from $P$ and $CP$ asymmetries in the background. The
$CPT$ breaking effect on the other hand, must necessarily arise from
the $CPT$ asymmetries in the background. Thus, for example, since the
weak interactions violate $P$ and $CP$ at some level, and since normal
matter is $CPT$-asymmetric, the {\em activity constant}\/ is present
in all normal matter.

The activity constant must also have a non-zero value in an
asymmetric background of neutrinos and antineutrinos.
In this case, parity is broken by the neutrino interactions. 
In addition, the required $CP$ and $CPT$ asymmetries arise
if the neutrino gas has a chemical potential, so that the number of 
neutrinos    
and antineutrinos in the background is not the same. For such a
system, $\Pi_P$ is then expected to depend on    
$n_\nu-n_{\bar\nu}$.

This effect can have in principle a number of
interesting physical implications.
For example, the standard big bang model
predicts a primordial neutrino background with a density comparable
to that of the microwave background radiation.
Light from all astrophysical
sources travel through this background before reaching us and,
if the background induces some optical activity,
it could have interesting physical effects on this light.

In this paper we present the 
calculation of the activity constant 
for a background of neutrinos and antineutrinos.
We do not assume any additional properties of the neutrinos beyond 
those of the standard electroweak model so that,
in particular, the neutrinos are assumed to be 
massless, two-component objects. In contrast to the
situations in which some physical effects depend
on unknown or conjectured properties of the neutrinos,
the implications that can be deduced
from the results of our calculations are based
on the well tested grounds
of the standard model.  While the possible
observational effects may be small in particular
physical situations, the effects are generally present at some level 
and can manifest themselves in unexpected circumstances.

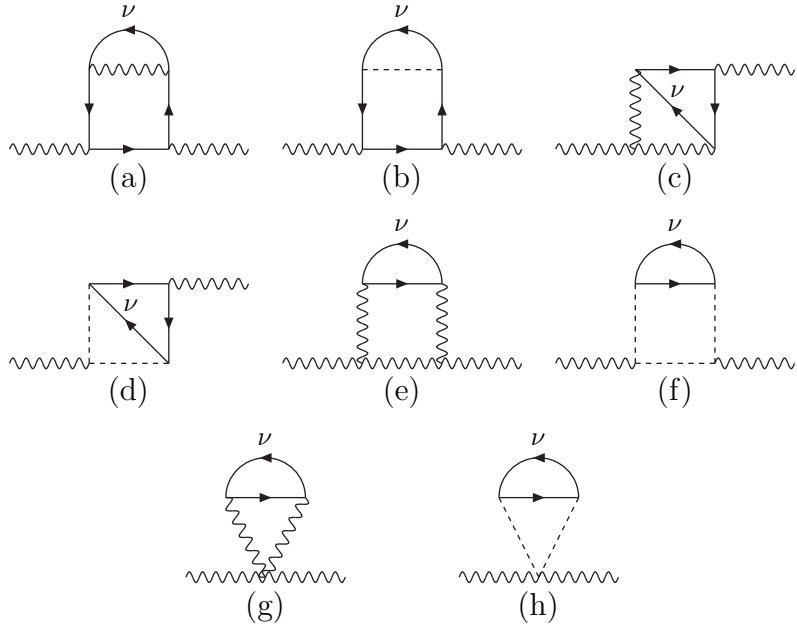
\begin{figure}
\begin{center}
\begin{picture}(100,60)(-5,0)
\ArrowLine(30,40)(30,10)
\ArrowLine(30,10)(60,10)
\ArrowLine(60,10)(60,40)
\ArrowArc(45,40)(15,0,180)
\Text(45,60)[b]{$\nu$}
\Photon(0,10)(30,10){2}{6}
\Photon(60,10)(90,10){2}{6}
\Photon(30,40)(60,40){2}{6}
\Text(45,5)[t]{\large (a)}
\end{picture}
\begin{picture}(100,60)(-5,0)
\ArrowLine(30,40)(30,10)
\ArrowLine(30,10)(60,10)
\ArrowLine(60,10)(60,40)
\ArrowArc(45,40)(15,0,180)
\Text(45,60)[b]{$\nu$}
\Photon(0,10)(30,10){2}{6}
\Photon(60,10)(90,10){2}{6}
\DashLine(30,40)(60,40){2}
\Text(45,5)[t]{\large (b)}
\end{picture}
\begin{picture}(100,60)(-5,0)
\ArrowLine(30,40)(60,40)
\ArrowLine(60,40)(60,10)
\ArrowLine(60,10)(30,40)
\Text(46,32)[]{$\nu$}
\Photon(0,10)(30,10){2}{6}
\Photon(60,40)(90,40){2}{6}
\Photon(30,40)(30,10){2}{6}
\Photon(30,10)(60,10){2}{6}
\Text(45,5)[t]{\large (c)}
\end{picture}
\begin{picture}(100,60)(-5,0)
\ArrowLine(30,40)(60,40)
\ArrowLine(60,40)(60,10)
\ArrowLine(60,10)(30,40)
\Text(46,32)[]{$\nu$}
\Photon(0,10)(30,10){2}{6}
\Photon(60,40)(90,40){2}{6}
\DashLine(30,40)(30,10){2}
\DashLine(30,10)(60,10){2}
\Text(45,5)[t]{\large (d)}
\end{picture} 
\begin{picture}(100,80)(-5,0)
\ArrowLine(30,40)(60,40)
\ArrowArc(45,40)(15,0,180)
\Text(45,60)[b]{$\nu$}
\Photon(30,40)(30,10){2}{6}
\Photon(30,10)(60,10){2}{6}
\Photon(60,10)(60,40){2}{6}
\Photon(0,10)(30,10){2}{6}
\Photon(60,10)(90,10){2}{6}
\Text(45,5)[t]{\large (e)}
\end{picture}
\begin{picture}(100,80)(-5,0)
\ArrowLine(30,40)(60,40)
\DashLine(30,40)(30,10){2}
\DashLine(30,10)(60,10){2}
\DashLine(60,10)(60,40){2}
\ArrowArc(45,40)(15,0,180)
\Text(45,60)[b]{$\nu$}
\Photon(0,10)(30,10){2}{6}
\Photon(60,10)(90,10){2}{6}
\Text(45,5)[t]{\large (f)}
\end{picture}
\begin{picture}(100,80)(-5,0)
\ArrowLine(30,40)(60,40)
\ArrowArc(45,40)(15,0,180)
\Text(45,60)[b]{$\nu$}
\Photon(30,40)(45,10){2}{6}
\Photon(60,40)(45,10){2}{6}
\Photon(15,10)(45,10){2}{6}
\Photon(45,10)(75,10){2}{6}
\Text(45,5)[t]{\large (g)}
\end{picture}
\begin{picture}(100,80)(-5,0)
\ArrowLine(30,40)(60,40)
\DashLine(30,40)(45,10){2}
\DashLine(60,40)(45,10){2}
\ArrowArc(45,40)(15,0,180)
\Text(45,60)[b]{$\nu$}
\Photon(15,10)(45,10){2}{6}
\Photon(45,10)(75,10){2}{6}
\Text(45,5)[t]{\large (h)}
\end{picture}
\end{center}
\caption[]{
\label{2ldiags}
Two-loop diagrams for the photon self-energy, with neutrinos
appearing as internal lines. The neutrino lines are marked as such. All other
internal fermion lines correspond to the charged lepton that couples to the
neutrino. The external gauge boson lines correspond to photons. 
All internal gauge boson lines are $W$-lines, and the internal scalar
lines denote the unphysical 
Higgs boson corresponding to the $W$-boson. 
In addition, there are crossed diagrams obtained by
interchanging the external photon lines.}
\end{figure}
\section{Strategies for the calculation}
\setcounter{equation}{0}
The basic technique for determining $\Pi_P$,
which has already been described in
Ref.~\cite{activity}, involves the calculation of the
photon self-energy diagrams involving internal neutrino lines
using the formalism of
Quantum Statistical Field Theory (more commonly and 
inappropriately called Finite Temperature Field Theory).
For the purpose of determining the neutrino background
contribution to $\Pi_P$ we take 
the background to consist only of neutrinos and antineutrinos,
and therefore we will adopt the propagator corresponding to the vacuum
for all the other particles.
For a background characterized by a temperature $1/\beta$ and chemical
potential $\mu$, the propagator for massless neutrinos is given by
	\begin{eqnarray}
iS_F(p,\beta,\mu) &=& \rlap/p \left[ {i\over p^2} - 2\pi\, \delta(p^2)
\eta_F(p,\beta,\mu) \right] \\*
&=& i S_0(p) + S'(p,\beta,\mu) \,,
\label{S}
	\end{eqnarray}
where $\eta_F$ contains the distribution functions for both neutrinos and 
antineutrinos:
	\begin{eqnarray}
\eta_F(p,\beta,\mu) = {\Theta (p\cdot u) \over e^{\beta(p \cdot u - \mu)} +1}
+ {\Theta (-p\cdot u) \over e^{-\beta(p \cdot u - \mu)} +1} \,.
	\end{eqnarray}
Here, $\Theta$ denotes the step function which has the value unity for
positive arguments, and is zero otherwise. Notice from Eq.\
(\ref{S}) that the propagator has two parts. The first one,
called $S_0$, is identical to the propagator in the vacuum. The
second part, which has been denoted by $S'$, is the correction
due to the background neutrinos, and is non-zero only for
on-shell neutrinos because of the factor $\delta(p^2)$.

Since the neutrinos do not couple to photons directly, the photon
self-energy diagrams involving internal neutrino lines occur
at the 2-loop level. These diagrams have been shown in
Fig.~\ref{2ldiags}. In drawing  these, we have adopted the
non-linear $R_\xi$ gauge \cite{nonlingauge} in which the
$Ww\gamma$ coupling vanishes, where $w$ is the unphysical Higgs
which is eaten up by the $W$-boson in the unitary gauge.

Notice that each diagram contains only one internal neutrino
line. Once we write down the amplitude corresponding to these
diagrams, there will be two kinds of terms. The first kind, which
will come from the vacuum part of the neutrino propagator, will
give a contribution to the vacuum polarization.  We are
interested in the other contribution which depends on the
background medium, involving the $S'$ term of the neutrino
propagator.

\begin{figure}
\begin{center}
\begin{picture}(100,50)(-5,0)
\ArrowLine(0,40)(30,40)
\ArrowLine(30,40)(30,10)
\ArrowLine(30,10)(60,10)
\ArrowLine(60,10)(60,40)
\ArrowLine(60,40)(90,40)
\Photon(0,10)(30,10){2}{6}
\Photon(60,10)(90,10){2}{6}
\Photon(30,40)(60,40){2}{6}
\Text(45,5)[t]{\large (a)}
\end{picture}
\begin{picture}(100,50)(-5,0)
\ArrowLine(0,40)(30,40)
\ArrowLine(30,40)(30,10)
\ArrowLine(30,10)(60,10)
\ArrowLine(60,10)(60,40)
\ArrowLine(60,40)(90,40)
\Photon(0,10)(30,10){2}{6}
\Photon(60,10)(90,10){2}{6}
\DashLine(30,40)(60,40){2}
\Text(45,5)[t]{\large (b)}
\end{picture}
\begin{picture}(100,50)(-5,0)
\ArrowLine(0,40)(30,40)
\ArrowLine(30,40)(60,40)
\ArrowLine(60,40)(60,10)
\ArrowLine(60,10)(90,10)
\Photon(0,10)(30,10){2}{6}
\Photon(60,40)(90,40){2}{6}
\Photon(30,40)(30,10){2}{6}
\Photon(30,10)(60,10){2}{6}
\Text(45,5)[t]{\large (c)}
\end{picture} 
\begin{picture}(100,50)(-5,0)
\ArrowLine(0,40)(30,40)
\ArrowLine(30,40)(60,40)
\ArrowLine(60,40)(60,10)
\ArrowLine(60,10)(90,10)
\Photon(0,10)(30,10){2}{6}
\Photon(60,40)(90,40){2}{6}
\DashLine(30,40)(30,10){2}
\DashLine(30,10)(60,10){2}
\Text(45,5)[t]{\large (d)}
\end{picture}
\\[5mm] 
\begin{picture}(100,50)(-5,0)
\ArrowLine(0,40)(30,40)
\ArrowLine(30,40)(60,40)
\ArrowLine(60,40)(90,40)
\Photon(30,40)(30,10){2}{6}
\Photon(30,10)(60,10){2}{6}
\Photon(60,10)(60,40){2}{6}
\Photon(0,10)(30,10){2}{6}
\Photon(60,10)(90,10){2}{6}
\Text(45,5)[t]{\large (e)}
\end{picture}
\begin{picture}(100,50)(-5,0)
\ArrowLine(0,40)(30,40)
\ArrowLine(30,40)(60,40)
\ArrowLine(60,40)(90,40)
\DashLine(30,40)(30,10){2}
\DashLine(30,10)(60,10){2}
\DashLine(60,10)(60,40){2}
\Photon(0,10)(30,10){2}{6}
\Photon(60,10)(90,10){2}{6}
\Text(45,5)[t]{\large (f)}
\end{picture}
\begin{picture}(100,50)(-5,0)
\ArrowLine(0,40)(30,40)
\ArrowLine(30,40)(60,40)
\ArrowLine(60,40)(90,40)
\Photon(30,40)(45,10){2}{6}
\Photon(60,40)(45,10){2}{6}
\Photon(15,10)(45,10){2}{6}
\Photon(45,10)(75,10){2}{6}
\Text(45,5)[t]{\large (g)}
\end{picture}
\begin{picture}(100,50)(-5,0)
\ArrowLine(0,40)(30,40)
\ArrowLine(30,40)(60,40)
\ArrowLine(60,40)(90,40)
\DashLine(30,40)(45,10){2}
\DashLine(60,40)(45,10){2}
\Photon(15,10)(45,10){2}{6}
\Photon(45,10)(75,10){2}{6}
\Text(45,5)[t]{\large (h)}
\end{picture}
\end{center}
\caption[]{
\label{Fdiags}
One-loop diagrams for neutrino-photon forward scattering,
obtained by cutting the neutrino line in the diagrams of
Fig.~\ref{2ldiags}. In these diagrams, the 
external fermion lines are
neutrinos. All other lines are as explained in the caption of
Fig.~\ref{2ldiags}. 
There are also crossed diagrams obtained by
interchanging the external photon lines.} 
\end{figure}
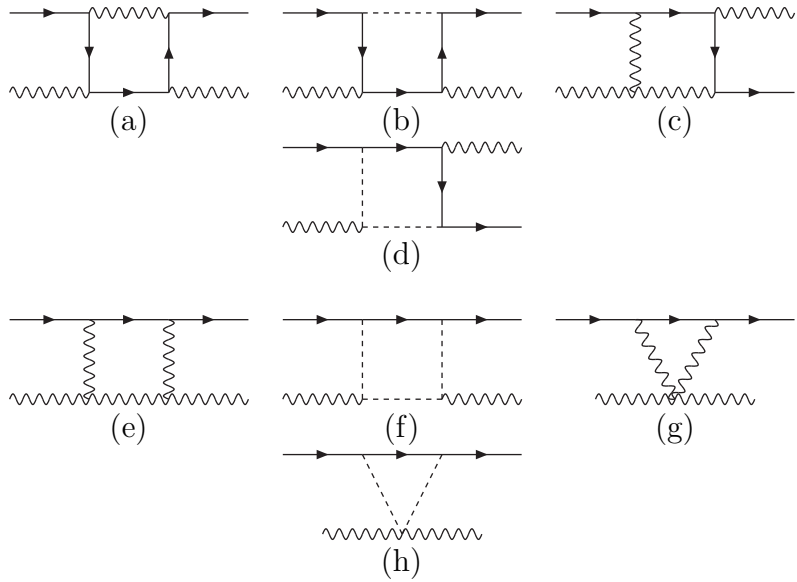

As we have already pointed out, the $S'$-part of the
propagator contributes only when the neutrino momentum $p$
satisfies the on-shell condition. If we cut the
neutrino line so that it becomes a pair of external lines,
then the resulting diagram contributes to the
neutrino-photon forward scattering amplitude
$M_{\mu\nu}(p,k)$. These diagrams are shown in
Fig.~\ref{Fdiags}. 
It then follows that, in 
the diagram where the neutrino line is
closed, the background-dependent  
contribution to the photon self-energy, denoted by
$\Pi'_{\mu\nu}(k)$, is given by
	\begin{eqnarray}
i\Pi^\prime_{\mu\nu}(k) = - \, \int {d^4p \over (2\pi)^4} \; \mbox{Tr}\,
\left[ S'(p) M_{\mu\nu} (p,k) \right] \,,
\label{intS'M}
	\end{eqnarray}
where the negative sign is due to the usual rule for a closed
fermionic loop.

Therefore, our first task is to calculate $M_{\mu\nu}(p,k)$. For
the moment, let us consider the diagrams for the
non-forward scattering amplitude,
and denote the momenta of the two external photons by
$k$ and $k'$. Since the external lines in those diagrams correspond
to the neutrinos, 
the photon is attached to internal lines only.  As a result, the
contribution of these diagrams is well-defined (i.e.,
non-singular) in the limits $k \to 0$, $k' \to 0$. This, combined
with the gauge-invariance conditions
	\begin{eqnarray}
k^\mu M_{\mu\nu} = 0 \,, \qquad
k'^\nu M_{\mu\nu} = 0 \,,
	\end{eqnarray}
implies that the amplitude must be of order $kk'$. In other
words, the amplitude must be proportional to at least one power
of $k$ and one power of $k'$. Since $k = k'$ for the forward scattering
amplitude, we thus conclude that $M_{\mu\nu}(p,k)$ must be of
order $k^2$.

This fact simplifies our task for the following reason. Except for the
diagram in Fig.~\ref{Fdiags}a, in all the other diagrams that
contribute to the forward $\nu$-$\gamma$ scattering the terms that
have at least two powers of external momenta are of order $1/M_W^4$
and higher, which are not important for small values of $k$ compared
to $M_W$.  Those diagrams also have terms of order $1/M_W^2$, but they
contain only one power of external photon momentum and, as argued
above, by gauge invariance they will cancel out at the end.  On the
other hand the diagram in Fig.~\ref{Fdiags}a can give a term that
contains at least two powers of the photon momentum and at the same
time is not necessarily of order $1/M_W^4$.  Denoting by $m$ the mass
of the charged lepton in the loop, that term is of order $1/(M_W^2
m^2)$ for $|k^2| < m^2$.  Thus, in order to determine the leading
contribution (in powers of $M_W^2$) to the gauge invariant part of the
amplitude we need to calculate only the diagram shown in
Fig.~\ref{Fdiags}a, retaining from that only the terms of order
$1/M_W^2$ that contain at least two powers of $k$.  The result so
obtained is proportional to $1/(M_W^2 m^2)$ and has significance in
the kinematic regime $|k^2| < m^2$. For the kinematic regime
$|k^2|>m^2$, of course, we need to calculate all the diagrams shown in
Fig.~\ref{Fdiags}.

\section{Details of the calculation}
\setcounter{equation}{0}
For the purpose of calculation, it is again convenient to start
with the general process
	\begin{eqnarray}
\nu(p) + \gamma(k) \to \nu(p') + \gamma(k') 
	\end{eqnarray}
and put $k=k'$ at the end. The amplitude for this process coming
from Fig.~\ref{Fdiags}a can be written as 
	\begin{eqnarray}
M_{\mu\nu}^{(a)} &=& - \left( {ig \over \surd 2} \right)^2 (-ie)^2
\gamma_\lambda L \left\{ T_{\lambda\mu\nu} (k,k') +
T_{\lambda\nu\mu} (-k',-k) \right\} \,,
	\end{eqnarray}
where $L=(1-\gamma_5)/2$, and
	\begin{eqnarray}
T_{\lambda\mu\nu} (k,k') &=& \mbox{Tr} \int {d^4q \over (2\pi)^4}
\; L \left( {i\over \rlap/q + \rlap/k - m} \right) \gamma_\mu 
\left( {i\over \rlap/q - m} \right) \gamma_\nu 
\left( {i\over \rlap/q + \rlap/k' - m} \right) \gamma_\lambda
\nonumber\\
& \times & 
\left( {-i\over (q - p + k')^2 - M_W^2} \right) \,,
	\end{eqnarray}
$m$ being the mass of the charged lepton in the loop. 
In writing this, we have adopted the Feynman gauge for the
$W$ propagator, and we have used the relation
	\begin{eqnarray}
\gamma^\alpha L A \gamma_\alpha L = - \gamma^\alpha L \;
\mbox{Tr} \,(A\gamma_\alpha L) \,,
	\end{eqnarray}
which is valid for any $4\times4$ matrix $A$. Adopting a different
gauge for the $W$ propagator, as well as including the
diagram involving the exchange of the corresponding
unphysical Higgs particle, would result in extra contributions
of order $1/M_W^4$ which, as already argued, we are not keeping. 
After some manipulations we obtain
	\begin{eqnarray}
\lefteqn{T_{\lambda\mu\nu} (k,k') + T_{\lambda\nu\mu} (-k',-k) =}\nonumber\\
& &
\mbox{Tr} \int {d^4q \over (2\pi)^4}
\left( {1\over \rlap/q + \rlap/k - m} \right) \gamma_\mu 
\left( {1\over \rlap/q - m} \right) \gamma_\nu
\left( {1\over \rlap/q + \rlap/k' - m} \right) \gamma_\lambda 
\nonumber\\*
&\times & 
\left( {R\over (q + p + k)^2 - M_W^2} - 
{L\over (q - p + k')^2 - M_W^2} \right) \,.
	\end{eqnarray}
At this stage, we can put $k=k'$, and replace the denominators
appearing in the $W$
propagators by $q^2-M_W^2$, because the extra terms will
contribute only to order $1/M_W^4$. The rest
of the calculation is straightforward and, using the convention
	\begin{eqnarray}
\varepsilon_{0123} = -1 \,,
	\end{eqnarray}
we obtain
	\begin{eqnarray}
	\label{resultamplitude}
M_{\mu\nu} = {-ie^2g^2 \over 8\pi^2 M_W^2 m^2} \gamma_\lambda L
\; (ik^2 \varepsilon_{\mu\nu\lambda\rho} k^\rho) J(k^2) \,,
\label{defJ}
	\end{eqnarray}
where
	\begin{eqnarray}
J(k^2) = \int_0^1 dx \; {x(1-x) \over 1 - x(1-x)(k^2/m^2)} \,.
	\end{eqnarray}

Substituting Eq.\ (\ref{resultamplitude}) in Eq.\ (\ref{intS'M})
and using the background-dependent part of the neutrino
propagator from Eq.\ (\ref{S}) this gives
	\begin{eqnarray}
i\Pi'_{\mu\nu}(k) &=& {-ie^2g^2 \over 8\pi^2 M_W^2 m^2} 
\; (ik^2 \varepsilon_{\mu\nu\lambda\rho} k^\rho) J(k^2) I_\lambda
\,,
\label{JI}
	\end{eqnarray}
where 
	\begin{eqnarray}
I_\lambda = \int {d^4p \over (2\pi)^4} \; \mbox{Tr}\, \left( \rlap/p
\gamma_\lambda L \right) 2\pi\, \delta(p^2) \left[ {\Theta (p\cdot u)
\over e^{\beta(p \cdot u - \mu)} +1} 
+ {\Theta (-p\cdot u) \over e^{-\beta(p \cdot u - \mu)} +1} \right] \,.
	\end{eqnarray}
Carrying out the trace, and performing the integration over $p_0$ 
in the rest frame of the medium with the help of the
$\delta$-function, we obtain 
	\begin{eqnarray}
I_\lambda = \left( n_\nu - n_{\bar\nu} \right) u_\lambda\,,
	\end{eqnarray}
where $n_{\nu,\overline\nu}$ stand for the number densities
of neutrinos and antineutrinos in the rest frame of the medium.
They are given by
	\begin{eqnarray}
n_{\nu,\overline\nu} = \int {d^3p \over (2\pi)^3}
\frac{1}{e^{\beta(|\vec p| \mp \mu)} +1} \,,
	\end{eqnarray}
with the upper (lower) sign holding for neutrinos
(antineutrinos), respectively.

This result can of course be used in any frame as long as we keep
in mind that the number densities of the neutrinos and antineutrinos
refer to the rest frame of the medium. Putting this back in Eq.\
(\ref{JI}) and comparing with the definition of $\Pi_P$ from
Eqs.\ (\ref{Pi}) and (\ref{PQR}), we finally obtain
	\begin{eqnarray}
\Pi_P^{(\nu)} (k) = {e^2 g^2 k^2 K\over 8\pi^2M_W^2 m^2} \left( n_\nu -
n_{\bar\nu} \right) J(k^2) \,,
     \label{Pip1}
	\end{eqnarray}
where $K$ was defined in Eq.\ (\ref{K}).  We have added a superscript
$\nu$ to $\Pi_P$ in Eq.\ (\ref{Pip1}) to remind us of the 
fact that this is the contribution to $\Pi_P$ from one neutrino specie
in the background (see below).

As for $J(k^2)$, we can neglect the $k^2$-dependent term in the
denominator if we are dealing with small $k^2$. The integration
then gives
	\begin{eqnarray}
J(k^2) = {1\over 6} \,.
\label{1/6}
	\end{eqnarray}
and Eq.\ (\ref{Pip1}) reduces to
	\begin{eqnarray}
\Pi_P^{(\nu)} = {\sqrt2 \,G_F\alpha  \over 3 \pi} \left({k^2 \over m^2}\right)
\left( n_\nu -n_{\bar\nu} \right)K 
\label{Pip2}
	\end{eqnarray}
So from far we have assumed that the background contains neutrinos
of only one flavor, which we have denoted generically
by $\nu$.  In general the background contains neutrinos
of various flavors, and each one gives a contribution
to $\Pi_P$ which is of the form given in Eq.\ (\ref{Pip2}).
Notice however that this result for 
$\Pi_P^{(\nu)}$ depends on the mass $m$ of the
charged fermion in the loop. Since the neutrinos
do not have any mass or mixing in the standard model, 
the internal charged lepton is the
electron if the thermal neutrino line corresponds to $\nu_e$, or either
the muon or the tau
if the thermal neutrino line corresponds to $\nu_\mu$ or
$\nu_\tau$, respectively.  We thus see that
the contribution to $\Pi_P$ coming from the asymmetries of
$\nu_\mu$ and $\nu_\tau$ are negligible since they are inversely proportional
to the corresponding charged lepton mass. Therefore, for
all practical purposes 
\begin{equation}
\label{Pip}
\Pi_P = {\sqrt2 \,G_F\alpha  \over 3 \pi} \left({k^2 \over m_e^2}\right)
\left( n_{\nu_e} -n_{\bar\nu_e} \right)K
\end{equation}

As already argued, the formula derived above is useful in the
kinematic regime $|k^2| < m^2$ because it gives the leading
contribution to $\Pi_P$, of order $1/M_W^2$, in
that regime.  On the other hand, it is not valid 
for $|k^2| > m^2$ because terms of the same order ($k/M_W^2$),
which eventually all cancel, have been omitted.
Therefore, a valid calculation for $|k^2| > m^2$
must include all the diagrams shown in Fig.~\ref{Fdiags}.
We have in fact performed that, and the result
of that task is presented in the rest of this section for the
sake of completeness. The reader interested in the physical
implications of our results may skip this part and go to the next
section.

For the purpose of compactness of notation, let us denote the
amplitude of the diagram $A$ in the form
	\begin{eqnarray}
\Pi'_{\mu\nu} {}^{(A)} = 3g^2e^2 (n_\nu -
n_{\bar\nu}) i \varepsilon_{\mu\nu\lambda\rho} u^\lambda k^\rho 
\int_0^1 dx\, \int_0^{1-x} dy\, \int {d^4q \over (2\pi)^4}
F^{(A)} \,.
	\end{eqnarray}
The task is now to present the quantities $F^{(A)}$ for each
diagram. We set the gauge parameter $\xi$ of the non-linear $R_\xi$
gauge to be equal to unity. The complete Feynman rules for this
choice can be obtained in Ref.~\cite{NPU}. Using these, it is
easy to see that diagrams \ref{Fdiags}f--h give no contribution
at all. For the other ones, we obtain
	\begin{eqnarray}
F^{(a)} &=& {x \left\{ 2(1-3y) q^2 + 4(1+y) m^2 + 4y^2(1-y)k^2
\right\} \over \left(q^2 - A^2\right)^4} \,, \nonumber\\
F^{(b)} &=& - 
{r x \left\{ (1-3y) q^2 + 2(1+y) m^2 + 2y^2(1-y)k^2
\right\} \over \left(q^2 - A^2\right)^4} \,, \nonumber\\
F^{(c)} &=& \int_0^{1-x-y} dz \; 
{ \left\{ -2q^2 + 8m^2 + 8 (x+z)(1-x-z) k^2
\right\} \over \left(q^2 - B^2\right)^4} \,, \nonumber\\
F^{(d)} &=& - 
{rq^2 \over \left(q^2 - B^2\right)^4} \,, \nonumber\\
F^{(e)} &=& 
{8x(q^2+yk^2) \over \left(q^2 - C^2\right)^4} \,,
	\end{eqnarray}
where
	\begin{eqnarray}
r=(m/M_W)^2 \,,
	\end{eqnarray}
and
	\begin{eqnarray}
A^2 &=& (1-x-y) M_W^2 + (x+y) m^2 - y(1-y) k^2 \,, \nonumber\\
B^2 &=& (1-x-y) M_W^2 + (x+y) m^2 - (x+z)(1-x-z) k^2 \,,
\nonumber\\
C^2 &=& (x+y) M_W^2 + (1-x-y) m^2 - y(1-y) k^2 \,.
	\end{eqnarray}
It is straight forward (though tedious) to show that if $k^2=0$,
the sum of all the diagrams vanish. This is in accordance with
the argument given earlier, showing that the forward scattering
amplitude must be at least of order $k^2$.

For small $k^2$, the integrals can be expanded in power series of
$k^2$, and the integrations of at least the $O(k^2)$-term can be
performed analytically. This gives, in the notation introduced in
Eq.\ (\ref{defJ}), 
	\begin{eqnarray}
J(k^2) & = & {1 \over 24(1-r)^3} \left[ 4 - 20r + 19r^2 -9r^3 -
{r\,\ln r \over 1-r} \left( 8 - 12r + 12r^2 -4r^3 \right)
\right]
\nonumber\\
& & \mbox{} 
+ O(k^2) \,,
	\end{eqnarray}
{}For small $r$, it is easy to see that this
reduces to the value given in Eq.\ (\ref{1/6}). 

\section{Dispersion relations and optical activity}
\setcounter{equation}{0}
The bilinear terms in the effective Lagrangian,
which describes the propagation of electromagnetic
waves through a medium, may be written in momentum
space in the form
	\begin{eqnarray}
{\cal L} = -\frac{1}{4}F^\ast_{\mu\nu}F^{\mu\nu} + {\cal L}'\,.
\label{one}
	\end{eqnarray}
The first term in Eq.\ (\ref{one}) is the standard kinetic energy term
for the electromagnetic field, while 
${\cal L}'$ is the contribution from the background medium, which
is given in terms of the photon self-energy by
	\begin{eqnarray}
\label{Lprime}
{\cal L}' = \frac{1}{2}A^{\ast\mu}\Pi_{\mu \nu}A^{\nu}\,.
	\end{eqnarray}

In physical applications we must include the effect of the 
charged plasma along with that of the neutrinos. 
While the main contribution to $\Pi_P$
is due to the neutrino background, as determined above,
$\Pi_{T,L}$ receive their main contribution from the
electron background. For our purposes, it is sufficient
to recall the well known result
	\begin{eqnarray}
\label{Pitl}
\Pi_{T,L}(\omega,K\rightarrow 0) = \omega^2_P \,,
	\end{eqnarray}
where $\omega^2_P$ is the plasma frequency.  Eq.\ (\ref{Pitl})
holds in the limit $K,\omega \ll m_e$ and it is useful
for finding the dispersion relations of the propagating modes
in the long wavelength limit. In general,
	\begin{eqnarray}
\label{omegap}
\omega^2_P = 4e^2\int\frac{d^3p}{(2\pi)^3 2E}
(f_e + f_{\overline e})\left[1 - \frac{p^2}{3E^2}
\right]\,,
	\end{eqnarray}
where $f_{e,{\overline e}}$ are the electron and positron
 distribution functions.
When the electron gas is in the non-relativistic limit, 
Eq.\ (\ref{omegap}) reduces to the standard formula
	\begin{eqnarray}
\omega^2_P = \frac{e^2n_e}{m_e}\,,
	\end{eqnarray}
where $n_e$ is the electron number density.
The equation of motion that follows from
Eqs.\ (\ref{one}) and (\ref{Lprime}) is
	\begin{eqnarray}
\label{eqmotion}
D^{-1}_{\mu\nu}A^\nu = 0\,,
	\end{eqnarray}
where
	\begin{eqnarray}
\label{invprop}
D^{-1}_{\mu\nu} = (-k^2 + \Pi_T)R_{\mu\nu} + (-k^2 + \Pi_L)Q_{\mu\nu}
+ \Pi_P P_{\mu\nu} \,.
	\end{eqnarray}

This equation may be written in terms of the components of $A^\mu$ by
means of the following device.  We introduce two vectors whose
components in the rest frame of the medium are $(0,\hat e_{1,2})$, and
where $\hat e_{1,2}$ are unit vectors orthogonal to $\vec k$ with the
convention that $\hat e_1, \hat e_2, \hat k$ form a right-handed
coordinate system.  Then, as shown in Ref. \cite{activity}, in any
frame these vectors satisfy
	\begin{eqnarray}
\label{e1e2relation}
e_{1\mu} = iP_{\mu\nu}e_2^\nu\,,\qquad e_{2\mu} = -iP_{\mu\nu}e_1^\nu \,.
	\end{eqnarray}
It is also useful to define 
	\begin{eqnarray}
\label{e3}
e^\mu_3 = \frac{\tilde u^\mu}{\sqrt{-\tilde u^2}} \,,
	\end{eqnarray}
which is orthogonal to $e_{1,2}^\mu$ and $k^\mu$.
If we now decompose $A^\mu$ in the form
	\begin{eqnarray}
A^\mu = A_1 e_1^\mu + A_2 e_2^\mu + A_L e_3^\mu + A_\parallel k^\mu 
	\end{eqnarray}
and substitute this in Eq.\ (\ref{eqmotion}), it follows
that the component $A_\parallel$ decouples from
the equations, as it should be since it is associated
with the gauge degree of freedom, while for the
longitudinal component we obtain the equation
	\begin{eqnarray}
\label{eqAL}
(-k^2 + \Pi_L) A_L = 0 \,,
	\end{eqnarray}
which implies the dispersion relation 
	\begin{eqnarray}
k^2 = \Pi_L \,.
	\end{eqnarray}
On the other hand, for the transverse modes
we obtain a set of coupled equations that can be written in
matrix form as
	\begin{eqnarray}
\left(\begin{array}{cc} 
-k^2 + \Pi_T & -i\Pi_P  \\
i\Pi_P & -k^2 + \Pi_T
\end{array}\right)
\left(\begin{array}{c}A_1\\ A_2 \end{array}
\right) = 0 \,.
\label{d2}
	\end{eqnarray}
In this way, the dispersion relation for the transverse modes is found
to be given by the solutions of the equations
	\begin{eqnarray}
k^2 = \Pi_T \pm \Pi_P = \omega_P^2 \pm \Pi_P
\label{dpm}
	\end{eqnarray}
with the corresponding polarization vectors
	\begin{eqnarray}
\label{e+-}
e^{(\pm)}_\mu = \frac{1}{\sqrt{2}}\left(e_{1\mu} \pm ie_{2\mu}\right) \,.
	\end{eqnarray}
Thus, the general solution for $A^\mu$ is of the form
	\begin{eqnarray}
A_\mu = A_{+}e^{(+)}_\mu + A_{-}e^{(-)}_\mu + A_L e_{3\mu} \,,
	\end{eqnarray}
with $A_{\pm}$ representing the amplitudes of the left($-$)
and right($+$) circularly polarized modes.

In this way we see that  the two transverse modes have 
different dispersion relations.
This is just the phenomenon of optical activity, which
in the present case is induced by the chiral nature of
the neutrino interactions. 
The dispersion relation for each mode can be obtained
explicitly by solving Eq.\ (\ref{dpm}). 
For this purpose, it is useful rewrite the expression for $\Pi_P$ as
	\begin{eqnarray}
\Pi_P = aR_\nu k^2 K \,,
\label{defa}
	\end{eqnarray}
where
\begin{equation}
R_\nu \equiv {n_{\nu_e} -n_{\bar\nu_e} \over n_0}
\label{Rnu}
\end{equation}
is the dimensionless form of the neutrino asymmetry, and
\begin{equation}
a \equiv  {\sqrt2 \,G_F\alpha  \over 3 \pi} \left({n_0 \over m_e^2}
\right)
\end{equation}
for any arbitrary benchmark value $n_0$. Taking $n_0=1\,{\rm
cm}^{-3}$, we obtain
	\begin{eqnarray}
a = 4 \times 10^{-43} \, {\rm GeV}^{-1}
= 8 \times 10^{-57} \, {\rm cm} \,.
	\end{eqnarray}
The solutions to Eq.\ (\ref{dpm}) can be obtained to the leading 
order in $a$ (i.e, leading order in $G_F$) 
by substituting $k^2 \simeq \omega_P^2$ 
on the right hand side of Eq. (\ref{defa}).  
In this way we find the dispersion relations
of the two modes to be given by
\begin{equation}
\label{omegapm1}
\omega_{\pm}(K) = \sqrt{K^2 + \omega_P^2} \pm \frac{1}{2}(aR_\nu
\omega_P^2)\frac{K}{\sqrt{K^2 + \omega_P^2}}\,.
\end{equation}
An alternative interpretation of these dispersion relations
is that, for a given value of $\omega$, 
the two circularly polarized modes travel with
different momenta 
	\begin{eqnarray}
K_\pm=(\omega^2- \omega_P^2)^{1/2}  \mp {1\over 2}\,  
a R_\nu \omega_P^2 \,.
\label{Kpm1}
	\end{eqnarray}
In any case, it follows that the two modes have different
phase and group velocities.
This fact can manifest itself in various ways, some of which we
discuss in the next section.

\section{Applications}
\setcounter{equation}{0}
\subsection{Differential time delay}
Consider a pulse of light originating at some
particular instant and subsequently 
propagating through the neutrino medium.  Since
the two circularly polarized components 
have different group velocities 
$v_g^\pm = \partial \omega_\pm/\partial K$, they
will disperse in time into two separate pulses. If the time delay
between them is larger than the width of the original
pulse, then it can be experimentally observable.  

From the dispersion relations in Eq.\ (\ref{omegapm1})
we obtain 
\begin{eqnarray}	
\label{vpm}
v_g^{\pm} & = & \frac{K}{\sqrt{K^2 + \omega_P^2}} \pm 
\frac{1}{2}aR_\nu \frac{\omega_P^4}{(K^2 + \omega_P^2)^{3/2}}
\nonumber\\
& \simeq & 
1 - \frac{\omega_P^2}{K^2} \pm  \frac{aR_\nu\omega_P^4}{2K^3}\,,
\end{eqnarray}
where, in the second formula, we have indicated
the approximate result in the limit $K \gg \omega_P$.  
At a distance $\ell$ from the emission point, the differential time delay
between the two pulses is given by
\begin{eqnarray}
\Delta t & = & \left| {\ell\over v_g^+}-{\ell\over v_g^-}\right| \nonumber\\  
\label{tpm}
& \simeq & {aR_\nu\omega_P^4 \ell\over K^3} \,,
\end{eqnarray}
where in the final step we have used the approximate
formula for $v_g^{\pm}$ given in Eq.\ (\ref{vpm}).

As an example of a realistic setting where these considerations
may be relevant, we mention the following.
Gamma ray burster (GRB) photons have an average fluence $F= 10^{-6}
\rm erg/cm^2$, per burst.  If they are at cosmological distances 
\cite{pac} $D=3000$~Mpc 
then the total energy output per burst $E=4\pi D^2F = 10^{51}$
ergs. 
Since this is equal to the gravitational binding energy of a typical
neutron star, they are thought to arise from the merger of neutron
stars. The gravitational energy of the neutron stars gets converted into
kinetic energy in the form of an expanding fireball. 
A temperature of the fireball of about $10$~MeV  explains the
observed spectrum of the signal. From the average  time variability of 
the burst which is 1\,---\,10~s, the size of the radiation zone is
expected 
to be $10^{11}$ to $10^{12}$\,cm. Assuming that the neutrinos are produced
by a  shift in the beta equilibrium of the neutron star, we can 
have $n_e \sim n_\nu \sim 10^3\;\rm MeV^3$. The
differential time delay between the right and the left polarized
components of a burst signal of mean energy 1~MeV, traveling through
this medium will have
a differential time lag $\Delta t \sim 10^{-6}$\,s, which may lie
within the observable range.
  
\subsection{Wavelength independent optical rotation}

As our second example, consider a monochromatic wave
propagating along a given direction, which we take
to be the $z$-axis.  If the wave is linearly polarized
(that is, it contains an equal admixture of the two
circular polarizations), then the amplitude of such a wave
is of the form
\begin{eqnarray}
\label{monoamp}
\vec A(z,t) & = & A_\omega e^{-i\omega t}\left(
e^{izK_{+}}\hat e^{(+)} + e^{izK_{-}}\hat e^{(-)}
\right)\nonumber\\
& = &  A_\omega e^{-i\omega t}e^{i\frac{1}{2}(K_{+} + K_{-})z}\left(
e^{-i\frac{1}{2}(K_{-} - K_{+})z}\hat e^{(+)} +
e^{i\frac{1}{2}(K_{-} - K_{+})z}\hat e^{(-)}
\right)\,.
\nonumber\\
\end{eqnarray}
We have implicitly chosen the origin and orientation of the
coordinate system in such a way that, at $z = 0$, the linear
polarization vector of the wave points along $\hat e_1$,
or the $x$-direction.
Using the relations $\hat e^{(\pm)} = (\hat e_1 \pm i\hat e_2)/\sqrt{2}$
implied by Eq.\ (\ref{e+-}), it follows that at any given distance
$z = \ell$ the polarization vector of the wave
points at an angle given by
\begin{eqnarray}
\label{phi}
\phi(\ell) = \frac{1}{2}(K_{-} - K_{+})\ell 
= \frac{1}{2}a R_\nu \omega_P^2 \ell 
\end{eqnarray}
relative to the $x$-axis.

It is known that such a rotation of the polarization vector
is also caused by the presence of an  external magnetic field.  This
effect, known as the Faraday rotation, is commonly used for measuring
extragalactic magnetic fields. The difference between the optical
rotation caused by neutrinos and the Faraday rotation
effect is that the
expression for $\phi$ given in Eq.\ (\ref{phi}) is 
independent of the wavelength $\lambda$ of the
propagating waves, whereas in the case of the Faraday rotation the 
angle is proportional to $\lambda^2$. 

For comparison with other sources of optical activity, it is 
useful to note that the contribution from the neutrino
background to the effective Lagrangian
given in Eq.\ (\ref{Lprime}) can
be written in the form
\begin{eqnarray}
\label{Lsubp}
{\cal L}_P = \frac{1}{2}A^{\ast\mu}S^\nu\tilde F_{\mu \nu} \,,
\label{lprime}
\end{eqnarray}
where $\tilde F^{\mu\nu} \equiv (1/2)\epsilon^{\mu \nu \alpha
\beta}F_{\alpha \beta}$ is the dual electromagnetic tensor and 
\begin{eqnarray} 
\label{Smu}
S^{\mu} & = & \frac{\Pi_P}{K} \, u^\mu\nonumber\\
&  = & a R_\nu k^2 u^\mu 
\simeq a R_\nu \omega_P^2 u^\mu \,.
\end{eqnarray}
In the second line of Eq.\ (\ref{Smu}) we have
used Eq.\ (\ref{defa}), and we have also put $k^2\simeq
\omega_P^2$ which is correct to leading order in $G_F$. 
It has been pointed out in the
literature that such an interaction breaks the invariance
under the $CPT$ transformation \cite{NP2}, and also breaks
Lorentz invariance \cite{cfj,gc} if $S_\mu$ is a fixed 4-vector (i.e.,
$S_\mu$ does not transform as a four-vector under Lorentz
transformations). 
As we have indicated above, one distinctive feature 
of this term is a rotation of the
polarization of electromagnetic waves by an
angle which is identical for all wavelengths.  
Using the observations about the 
polarization of the radiation from distant sources, 
various limits have been deduced on possible deviations from the
physics of the standard model \cite{cfj,hs,mn1,mn2}. In addition,
it has been recently 
claimed \cite{rn} that the effect induced by a term of the kind
given in Eq.\ (\ref{Lsubp}) is actually observed in the data, 
indicating that the radiation from radio 
galaxies is polarized along a preferential axis with respect to the
galaxy \cite{rg}. This claim has been contested by several
authors \cite{cf}, and it has been argued that the data
should instead be
interpreted as an upper bound on the possible non-magnetic optical
rotation.  

We want to point out here that the effect of a
wavelength-independent optical rotation occurs even
in the standard model of particle interactions, 
if the universe has an
unequal number of neutrinos and antineutrinos.  Moreover,
it is present even if the neutrinos are
strictly massless. This was pointed out earlier in a qualitative
manner in Ref.\ \cite{activity}. The new input of
the present work is to specify that the magnitude of $S_\mu$ in such a 
case is given by Eq. (\ref{Smu}). An estimate of
it can be obtained as follows.

If the cosmological neutrino background has a
non-zero chemical potential, then it is bounded by the constraint that
the neutrino energy density should not exceed the closure density of
the universe, $\mu_{\nu}/8 \pi^2 < 2 \times 10^{-47} \, {\rm GeV}^4$,
which means that $R_\nu =\mu_\nu^3/(6 \pi^2 n_0) \leq 5 \times 10^5$.
With $\omega_P^2 = 4 \pi \alpha
n_e/m_e$, taking $n_e = 0.03 \, {\rm cm}^{-3}= 2.3 \times 10^{-43} \,
{\rm GeV}^3$ in the intergalactic medium and $\ell=H_0^{-1} = 0.5
\times 10^{42} \, {\rm GeV}$ the magnitude of the optical rotation
given by Eq.\ (\ref{phi}) is of order $10^{-36}$. 
This seems to be too small to be observable.
Therefore, if the observation of a wavelength-independent optical
rotation is firmly established, then it probably 
cannot be due to the cosmological
background neutrinos, at least within the assumptions of
the standard particle physics and cosmological models.

\subsection{Differential bending of light}
Consider the bending of light by a massive body in a neutrino
medium. This could occur in the halo of a galaxy, in the
exterior of a neutron star or the neutrino-sphere of a
supernova. In empty space, a light ray bends 
by an angle $\theta = 4 GM/b$, where $M$ is the mass of the body
and $b$ is the impact parameter. On the other hand
it has been shown that the gravitational bending angle
for a massive spinless particles depends on the mass
of the particle \cite{ggp}. This suggests that the bending angle
for a photon propagating in a matter background depends on the
dispersion relation of the photon. In fact, it has been found
that interactions 
of the type given in Eq.\ (\ref{Lsubp}) indeed give 
polarization-dependent bending for light \cite{mn1,mn2}.

Thus, when light propagates in a neutrino medium, the
left and the right polarized components bend through different
angles, which we denote by $\theta_\pm$. 
The image of the source splits into
two images, separated by an angle $\Delta \theta = \theta_+ -
\theta_-$.
To calculate the bending angle  we use the method given in
Ref.~\cite{oha}, adapting it to the present case
in which $\omega \neq K$.
The starting point is the equation of motion for a light wave of
momentum $k^\mu$ which, in the limit of geometrical optics, is
	\begin{eqnarray}
dk_\mu = {1\over 2} (\partial_\mu h^{\alpha}{}_\beta) k_\alpha
dx^\beta \,,
\label{dkmu}
	\end{eqnarray}
where 
	\begin{eqnarray}
h_{\alpha\beta} = g_{\alpha\beta} - \eta_{\alpha\beta} \,,
	\end{eqnarray}
with $\eta_{\alpha\beta}$ being the flat-space metric. As in
Ref.~\cite{oha}, we consider a ray of light with a given
polarization $\lambda = \pm$, passing by a
gravitational body of mass $M$ with impact parameter $b$. Taking
the $z$-axis along the direction of incidence, we then have
	\begin{eqnarray}
dx^\beta &=& \left( {1\over v_\lambda}, 0,0, 1 \right) dz \nonumber\\
k^\alpha &=& \left( 1, 0,0, v_\lambda \right) \omega \,,
\label{kdx}
	\end{eqnarray}
where $v_\lambda = K_\lambda(\omega)/\omega$. 
The bending angle is  given by
	\begin{eqnarray}
\theta_\lambda = {\Delta k_x \over k_z} \,,
\label{bendang}
	\end{eqnarray}
where $\Delta k_x$ is the total change in $k_x$. Using the
expression for $h_{\alpha\beta}$ corresponding to the
Schwarschild metric and using Eqs.~(\ref{dkmu}) and (\ref{kdx}),
we obtain
	\begin{eqnarray}
\Delta k_x = {GMb\omega \over v_\lambda} (1+v_\lambda^2) 
\int_{-\infty}^\infty dz
\; {1\over (z^2+b^2)^{3/2}} = {2GM\omega \over bv_\lambda} (1+v_\lambda^2) \,,
	\end{eqnarray}
and from Eq.~(\ref{bendang})
	\begin{equation}
\theta_\lambda = {2GM \over b} \left( 1+{1\over v_\lambda^2} \right) \,.
	\end{equation}
Therefore, the bending angles for the two polarizations
differ by an amount
\begin{equation}
\Delta \theta \equiv \theta_+ - \theta_-  =  {2GM \over b} \left({1\over
v_+^2} - {1\over v_-^2} \right) \,.
\end{equation}
Remembering that $v_{\pm} = K_{\pm}(\omega)\omega$ and using
Eq.\ (\ref{Kpm1}), we then find that
\begin{equation}
\Delta\theta \simeq  {4GM \over b} {a R_\nu \omega_P^2 \omega^2
\over (\omega^2-\omega_P^2)^{3/2}}
\end{equation}
for the neutrino background.

\section{Conclusions} 
\setcounter{equation}{0}
A background in which the number of neutrinos and antineutrinos
are not the same is an optically active medium.
This observation, first pointed out in Ref.\ \cite{activity},
has several implications and could in principle
lead to potentially important applications.
In this paper we have presented explicitly the results
of the calculation of the neutrino background contribution
to the photon self-energy in such a medium.  This has
allowed us to deduce the photon dispersion relations
for the two circularly polarized states of the 
photon traversing the neutrino background, and analyze
in some detail several  of the possible physical
effects.  We have shown, in particular, that the neutrino
background induces a difference
in the time delay between the light signals of opposite
polarizations, which could lie in an observable range
for photons originating in gamma ray bursters.
We have  also shown that the optical rotation effect
caused by the neutrinos is independent of the
wavelength of the electromagnetic wave, 
in contrast with the Faraday effect caused
by the presence of a magnetic field, which depends
on the square of the wavelength.  In addition,
we derived the formula for the bending angle of a light
ray passing by a massive object embedded in the
neutrino background, taking into account that the photon
dispersion relation is not the same as in the vacuum,
and that it is in fact different for each polarization mode.

At first sight, all these effects tend to be small. However,
we point out once more that the calculations that
we have described, and the results that we have obtained,
do not depend on any physical assumptions beyond those
required by the standard model of particle interactions,
including the question of whether or not the neutrinos
have a non-zero mass.  Hence, the 
effects that we have considered are present at some level.
It would not be the first time that a seemingly unimportant
feature of the electromagnetic interactions of neutrinos
in a background medium, such as the neutrino index of refraction
of a neutrino in the presence of a magnetic field \cite{DNP}, finds
its way much later to become the focus
of an important problem \cite{ks}. In the presents case, we have
considered several consequences related to the optically
active (chiral) nature of a neutrino background,
some of which could be observationally testable, 
and should in any case be kept in mind.

This work have been partially supported by the US National
Science Foundation Grant PHY-9600924 (J.F.N).

\end{document}